\documentclass[12pt,dvips]{article} \usepackage[dvips]{epsfig}
\newcommand{\be}{\begin{equation}}
\newcommand{\ee}{\end{equation}}
\newcommand{\bea}{\begin{eqnarray}}
\newcommand{\eea}{\end{eqnarray}}
\newcommand{\NC}{N_c}
\begin{document}
\title{
\vskip -4.5cm
\begin{flushleft}
{\small BUHEP-01-33; 
IFT-UAM/CSIC 01-37; CPT-2001/PE.4288}
\end{flushleft}
\vskip 1.5cm
Fermions on tori in uniform abelian fields
}
\author{L.~Giusti${}^{(a,b)}$, A.~Gonz\'alez-Arroyo${}^{(b,c)}$,
Ch.~Hoelbling${}^{(b,f)}$,  \\
H. ~Neuberger${}^{(d,e)}$  and  C.~Rebbi${}^{(b)}$\\
{${}^{(a)}$ \small Centre de Physique Theorique}\\
{\small CNRS Luminy, Case 907}\\
{\small F-13288 Marseille Cedex 9, France}\\
{${}^{(b)}$ \small Boston University Physics Department}\\
{\small 590 Commonwealth Avenue}\\
{\small Boston MA 02215, USA} \\
{${}^{(c)}$ \small Instituto de F\'{\i}sica Te\'orica
UAM/CSIC$^{*}$}\\
{\small Univ. Aut\'onoma de Madrid}\\
{\small Cantoblanco, Madrid 28049, SPAIN } \\
{${}^{(d)}$ \small LPTHE}\\
{\small Univ. Paris VI, T.16,  4 Place Jussieu, $1^{\rm er}$ etage,}\\
{\small 75252 Paris Cedex 05, France} \\
{${}^{(e)}$ \small Department of Physics and Astronomy\footnote{Permanent Address.}}\\
{\small Rutgers University}\\
{\small Piscataway, NJ 08855, USA} \\
{${}^{(f)}$ \small NIC/DESY Zeuthen}\\
{\small Platanenallee 6 }\\
{\small D-15738 Zeuthen, Germany} 
}
\maketitle
\abstract{We study Fermi fields defined on tori in the presence of
gauge backgrounds carrying non-trivial topology. 
We show that $2k$ dimensional field space can alternatively be
described by fields over a $k$-dimensional space.
This dual description is particularly
natural when the background is uniform and abelian.
The reduction in number of dimensions carries over
to the lattice. The lattice ultraviolet regularization induces an infrared
regularization of the lower dimensional representations.  We focus on $k=1,2$.}

\section{Introduction}

Dirac fields defined over even dimensional
tori can, for certain twists in
the gauge background, be equivalently represented by fields
defined over infinite flat spaces of half the number of dimensions.
The two descriptions are related by a change of basis. The reduction
in dimensionality comes from using Fourier space for half of the space-time
variables. The integral Fourier momenta are used to string the remaining
segments of continuous directions into infinite lines. This works for sections
of non-trivial bundles. 
Although
the two space-times are distinct in the topological sense, the
fermionic index stays the same. For simple backgrounds, consisting of 
a constant abelian field strength, the field equations are local
in either dimensionality. Extra interactions would be required
to single out one of the alternative descriptions as more natural.

The setup we are looking at has a long history. It has been
analyzed before in various contexts~\cite{Landau, abrikosov, 
saviddy, leutwyler, thooft, vanbaal2, vanbaal1}. Early studies in 
lattice field theory were carried out in~\cite{smitvink}.
Probably the main objection to developing a lattice version
was the absence of an exact fermionic ``viewpoint'' of background topology.
Fermions react to topological features in the background in
ways strongly correlated to their chiral properties and without
a lattice version of chirality only an approximate analysis was 
possible. This has changed with the domain wall formulation
of lattice fermions~\cite{domain} and, especially, with the overlap 
formulation~\cite{lattchiral}. Among recent studies on the response
of fermionic fields to topology in the overlap regularization we 
mention~\cite{lattchiral,gattringer1,chiu,gattringer2}.

In this paper we focus mainly on the aspect of alternate description
in terms of reduced dimensions. In the continuum, for our special
backgrounds, locality is preserved and the equations of motion 
stay differential in either description.\footnote{The description in
terms of lower dimension fields is not restricted to Dirac fields:
it would also be useful for analyzing the stability of the gauge
backgrounds against small perturbations in the gauge fields
~\cite{vanbaal1,gpgap}.} 
Our new contribution in this paper is to
provide lattice analogues to the reduced dimension description.
For example, we learn from this
that the spectrum of lattice fermions moving on an $L_1 \times L_2 $
toroidal lattice in a uniform abelian gauge background 
depends only on the area $L_1 L_2$. There is no
distinction between the arbitrary $L_1 \times L_2$ case and that
on a degenerate lattice of dimensions $L_1 L_2 \times 1$.

In sections 2 and 3 we introduce the continuum version of the representations
in two and four dimensions respectively. The origin of the dimensional
reduction is traced  back to a natural  Heisenberg algebra defined in this
system.
This algebra plays a central role in relating  the developments of
references \cite{tek,sushel} to the the general area  of Non-Commutative Field 
Theories, as reviewed, for example, in \cite{douglas}.                     
In section 4 we focus on 
the Dirac equation in constant abelian backgrounds, still in the continuum.
Sections 5 and 6 deal with the lattice version in two and four dimensions
respectively. We collect some numerical and asymptotic results in section 7.

\section{Two dimensions -- continuum}
\label{continuum}
We consider a two dimensional torus 
of size $l_1\times l_2$.  This can be viewed as the plane 
${\mathbf R}^2$ modulo the lattice 
generated by $e_1$, $e_2$, chosen for 
simplicity as $e_1=(l_1,0)$ and $e_2=(0,l_2)$.
The fields 
$\psi(x)\equiv\psi(x_1,x_2)$ are 
complex fields satisfying certain periodicity properties. 
Our fields are coupled to a gauge field and are thus sections of a $U(1)$ bundle
on the 2-torus. The bundle can be twisted, the twist being characterized by the
first Chern class $c_1$.  The first Chern number gives the 
topological charge: it measures the flux through
the two torus in units of $2 \pi$ (the charge $e$ is set to $1$).  

Periodicity of  fermion fields has to be taken modulo gauge transformations:
\bea
\psi(x+e_1)&=&\Omega_1(x)\,\psi(x)\label{eq11}\\
\psi(x+e_2)&=&\Omega_2(x)\,\psi(x) \label{eq12} 
\eea
The $U(1)$ fields $\Omega_i(x)$ are the transition functions made consistent by
\be
\label{consistency}
\Omega_1(x+e_2)\, \Omega_2(x)= \Omega_2(x+e_1)\, \Omega_1(x)
\ee  
By gauge transformations the 
transition functions are brought to the form 
\bea
\label{BConds}
\Omega_1(x)&=&\exp\{\imath \pi q x_2/l_2\} \\
\Omega_2(x)&=&\exp\{-\imath \pi q x_1/l_1\}
\eea 
The consistency
condition~(\ref{consistency}) forces $q$ to take only
integer values. $q$ is the first Chern number (often referred to
as winding number, in this set-up). 

Fields satisfying the boundary conditions (\ref{eq11}, \ref{eq12})make up
a Hilbert space ${\cal H}_q$ with scalar product
\be
\langle\phi|\psi\rangle= \int_{T^2}dx\; \phi^*(x)\psi(x)
\ee 
The integration is over the torus and because of the
integrand's periodicity can be performed over any  fundamental cell.

${\cal H}_q$ is left invariant by a ring of 
linear transformations generated by two unitary 
transformations ${\mathbf U}_i$.
\be
{\mathbf U}_i\psi(x)= \exp\{2 \pi \imath x_i/l_i\}  \psi(x)
\ee

Ordinary
translations map out of ${\cal H}_q$ because the transition functions
depend on $x$. This can be compensated by an appropriate gauge 
transformation leading to modified operations of translation by $a_i$:
\bea  
{\cal T}_1(a_1)\Psi(x)&=&\exp\{\imath \pi q a_1 x_2/{\cal A}\}\ \Psi(x+(a_1,0))\\
{\cal T}_2(a_2)\Psi(x)&=&\exp\{-\imath \pi q a_2 x_1/{\cal A}\}\
\Psi(x+(0,a_2))
\eea
where ${\cal A}=l_1 l_2$ is the area of the torus. The new 
translations along the two axes do not commute:
\be
{\cal T}_1(a_1){\cal T}_2(a_2)=\exp\{ -2\pi \imath q a_1 a_2/{\cal A}\}\ 
{\cal T}_2(a_2){\cal T}_1(a_1)
\ee
Elements of the
non-commutative group of translations do not commute with 
the discrete abelian group of 
transformations generated by ${\mathbf U}_i$. 
There exists only a finite subgroup of unitary transformations, generated
by $K^{(1),(2)}$ which commutes with translations: 
\bea
K^{(1)}&=&{\mathbf U}_1\  {\cal T}_2(l_2/q)\\
K^{(2)}&=&{\mathbf U}_2\  {\cal T}_1(-l_1/q)
\eea 
The boundary conditions imply  $(K^{(\mu)})^q={\mathbf I}$ and 
the operators $K^{(\mu)}$ obey:
\be
K^{(1)} K^{(2)}= \exp\{ 2 \pi \imath/q   \}\  K^{(2)} K^{(1)}
\ee
Any one of the $K^{(\mu)}$ generates a $Z_q$ group under which 
${\cal H}_q$ decomposes into $q$ orthogonal subspaces:
\be
 {\cal H}_q = \oplus_{n=1}^q  {\cal H}_{q, n}
\ee
Decomposing under $K^{(2)}$, fermion fields in 
${\cal H}_{q, n}$ satisfy:
\be
K^{(2)}\psi(x)=\exp\{-2 \pi \imath\frac{n}{q}\}\; \psi(x)
\ee
$K^{(1)}$ maps these subspaces into each other, but
the non-commutative translations leave the subspaces invariant.

The generators of translations along each 
axis are given by:
\be
D^{(0)}_i=\partial_i + \pi \imath q \epsilon_{ i j} x_j /{\cal A}
\ee
where $\epsilon_{ i j}$ is the antisymmetric tensor with two 
indices($\epsilon_{1 2}=1$).
The generators coincide  with ordinary covariant derivatives in a constant 
magnetic field:
\bea
A_i(x)=-\pi  q \epsilon_{ i j} x_j /{\cal A}\\
F=\partial_1A_2-\partial_2A_1=\frac{ 2 \pi  q}{{\cal A}}
\eea
The non-commutative translation group is made out of 
parallel transporters in this field. $q$ is 
given by the total flux.
The anti-self-adjoint generators of parallel transport do not commute:
\be
\label{HAlgb}
[D_1^{(0)},D_2^{(0)}]=-\imath F=\frac{ -2 \pi \imath q}{{\cal A}} 
\ee
Up to a rescaling, we have a Heisenberg algebra associated with one degree
of freedom and we are led
to a quantum mechanical view of the space as a Hilbert space
of a particle in  one-dimension. For this particle, 
the covariant derivatives play the role 
of position and momentum operators.
For $q > 0$, we take 
$ {\mathbf P} \leftrightarrow -\imath D_2^{(0)}$ and  
${\mathbf Q} \leftrightarrow -\imath D_1^{(0)}/F$ producing a traditional
position space representation in terms
of functions of a single real variable $h(z)$. 
The space  ${\cal H}_q$ is identified with  $\oplus_q L_2({\mathbf R})$ with
the $D_i^{(0)}$ action given by:
\bea
D_1^{(0)} h_n(z)= i F z\, h_n(z)\\
D_2^{(0)} h_n(z)=  \frac{dh_n(z)}{dz}
\eea

The fields get mapped into the quantum mechanical {\em position space} basis as follows:
Since the function
\be
\varphi(x_1,x_2)= \exp\{-\imath \pi q\frac{x_1 x_2}{{\cal A}}\}\ \psi(x_1,x_2) 
\ee
is periodic in the variable $x_1$ with period $l_1$ we can expand
it in a discrete Fourier series:
\be
\varphi(x_1,x_2)= \sum_k \exp\{2 \pi \imath\frac{x_1}{l_1}k\}\
h_k(x_2) 
\ee 
The periodicity requirement in $x_2$ imposes a constraint on
$h_k(x_2)$:
\be
h_{k+q}(x_2)=h_k(x_2+l_2)  
\ee
The constraint is solved by:
\be
h_k(x_2)=h_n(x_2+k l_2/q)
\ee
with $n=k \bmod q$. The expression 
for the field becomes:
\begin{eqnarray}
\label{changebasis}
\psi(x_1,x_2)= 
\exp\{\imath \pi q\frac{x_1 x_2}{{\cal A}}\}\ \times & \nonumber \\
\sum_{n=1}^q \sum_{s\in{\mathbf Z}}\exp\{2 \pi \imath\frac{x_1}{l_1}(n+sq)\}\
& h_n(x_2+\frac{n+qs}{q}l_2)
\end{eqnarray}
The {\em wave-function} $h_n(z)$ transforms as
required under  $D_1^{(0)}$ and $K^{(2)}$ and the index $n$ labels
the subspace ${\cal H}_{q,n}$ to which the component belongs.

\section{Four dimensions -- continuum}

We consider now fields defined
over a four-dimensional torus of size $l_1\times l_2\times l_3 \times
l_4$. If the gauge group is $U(\NC)$ 
the bundles are classified by the
instanton number $Q$    and by the first Chern numbers $c_{\mu\nu}$ associated to the
individual planes. If the gauge group is $SU(\NC)$ we only have
the instanton number, but if all the fields are insensitive to the center
$Z(\NC)$ the true gauge group becomes $SU(\NC)/Z(\NC)$ 
and there is a remnant of the first Chern numbers, now defined modulo
$\NC$, labeling the so called twist sectors. Excluding the case
$\NC=2$ with zero twist and odd instanton number the topological character
can be encoded in transition functions $\Omega_{\mu} (x)$ 
restricted to the $U(1)^{\NC-1}$
subgroup of $SU(\NC)$~\cite{vanbaal1,review}. The matter fields obey the
boundary conditions:
\be
\psi(x+e^{(\mu)})=\Omega_{\mu}(x)\psi(x)
\ee
The (not necessarily orthogonal) 4-vectors   $e^{(\mu)}$ have lengths 
$l_{\mu}$ and span a non-degenerate four dimensional lattice
$\Lambda$. The dual lattice is generated by $\tilde{e}^{(\mu)}_{\nu}$.
Denoting by $R_{\mu\nu}=e^{(\mu)}_{\nu}$ 
the components of the invertible matrix $R$, $\tilde{e}^{(\mu)}_{\nu}=
(R^{-1})_{\nu\mu}$.

In the Abelian description we consider matter fields in the fundamental 
representation with
all $\NC$ color components satisfying decoupled boundary 
conditions:
\be
\label{fourdBC}
\psi^a(x+e^{(\mu)})=\Omega^{aa}_{\mu}(x)\psi^a(x)
\ee
We again can pick transition matrices of the form:
\be
\label{TrMat}
\Omega^{aa}_{\mu}(x)=\exp\{\imath \pi Q^a_{\mu \nu} \tilde{e}^{(\nu)}_{\rho}
x_{\rho} \}
\ee
The integers $Q^a_{\mu\nu}$ are antisymmetric in $\mu ,\nu$ and can be assembled
into the matrix $Q^a$. 
For $U(\NC)$ the 
topological invariants are:
\begin{eqnarray}
\label{topinva}
c_{\mu \nu}&=&\sum_a Q^a_{\mu \nu}\\
\label{topinvb}
Q&=&\sum_a \mbox{Pf}(Q^a) 
\end{eqnarray} 
$\mbox{Pf}$ stands for Pfaffian.  
For $SU(\NC)$, $c_{\mu \nu}=0$. 

We study the space of 
fields subject to the boundary conditions ~(\ref{fourdBC})-(\ref{TrMat}).
We deal with one of the components only and drop the
color index; the number $Q$ and the matrix $Q^a$ will
become distinguishable only by context. We assume that $\mbox{Pf}(Q)\ne 0$. 

We again analyze classes of unitary operators that act in this space, starting 
with operations of multiplication by periodic functions. The 
generators of this discrete infinite abelian group are:
\be
U^{(\mu)}=\exp\{2 \pi \imath \tilde{e}^{(\mu)}_{\nu} x_{\nu}\} 
\ee
A translation by $b$, suitably modified
to respect the boundary conditions, is given by:
\be
{\cal T}(b)\psi(x)= \exp\{\frac{\imath}{2}F^{(0)}_{\mu \nu} b_{\mu} x_{\nu}
\}\; \psi(x+b) 
\ee  
The antisymmetric matrix $F^{(0)}_{\mu \nu}$ is given by: 
\be
\label{Fdef}
F^{(0)}_{\mu \nu}= 2 \pi Q_{\rho \tau}\tilde{e}^{(\rho)}_{\mu}
\tilde{e}^{(\tau)}_{\nu}
\ee
Translations only 
commute up to a phase:
\be
{\cal T}(b) {\cal T}(d)=\exp\{\imath F^{(0)}_{\mu \nu} b_{\nu} d_{\mu}\}\
{\cal T}(d) {\cal T}(b)
\ee
As in the two-dimensional case, we modify the $U^{(\mu )}$'s to achieve
commutativity with translations.
The generators of the
resulting non-abelian group are given by
($Q$  is invertible since by assumption 
its determinant does not vanish):
 \be
 K^{(\mu)}= U^{(\mu)}{\cal T}(-(Q^{-1})_{\mu \nu} e^{(\nu)})
 \ee

They satisfy:
\be
\label{twisteaters}
K^{(\mu)} K^{(\nu)} =\exp\{-2 \pi \imath (Q^{-1})_{\mu \nu}\}\   K^{(\nu)}
K^{(\mu)}
\ee
The representations of this group, directly related to the
Heisenberg group, have been studied
in~\cite{te} (see \cite{review} for a review). 
The irreducible representations have dimensions $\mbox{Pf}(Q)$.
Our boundary conditions make 
the group not only finite, but also irreducible (elements commuting with 
all the generators are multiples of the identity) and 
the space of fields satisfying the boundary conditions (for a fixed
color component) decomposes into a direct sum of $\mbox{Pf}(Q)$ 
spaces. Translations do not mix these subspaces. As we shall see 
later, there is exactly one
zero-mode in each of these subspaces, giving a total of $Q$, as expected 
(\ref{topinvb}).
 
The infinitesimal generators of translations are given by
\be
D^{(0)}_{\mu}=\partial_{\mu}+ \frac{\imath}{2}\, F^{(0)}_{\mu \nu}
x_{\nu} 
\ee
that is, covariant derivatives in a gauge field of uniform
field strength:
\be
[D^{(0)}_{\mu}, D^{(0)}_{\nu}]= -\imath F^{(0)}_{\mu \nu}
\ee

Before we derive a parametrisation of the space in terms of fields living
in only two dimensions we exploit symmetries to simplify the problem.
By transformations in $SL(4,{\mathbf Z})$ we bring 
$Q_{\mu \nu}$ to a
canonical form with
$Q_{12}=-Q_{2 1}=q$ and $Q_{0 3}=-Q_{3 0}=q'$ and all other entries vanishing. 
The Pfaffian is equal to $q q'$.

We now perform space-time coordinate transformations:
\be
x_{\mu} \longrightarrow x'_{\mu}=l_{\mu}\, \tilde{e}^{(\mu)}_{\nu}
x_{\nu}
\ee
In general, the transformation is non-orthogonal 
changing the metric tensor and the spectrum of  the Dirac 
operator. However, the previous formulae remain valid, and  
the matrix $F^{(0)}_{\mu \nu}$ is simpler now, with only
$F^{(0)}_{0 3}=-F^{(0)}_{ 3 0}= F'=2 \pi q'/(l_0 l_3)$ 
and $F^{(0)}_{12}=-F^{(0)}_{2 1}= F=2 \pi q/(l_1 l_2)$ as non-zero entries.
As a result, the algebra of covariant derivatives decouples, 
becoming the Heisenberg algebra for a quantum particle in two dimensions
and the problem factorises over the two planes $1-2$ and $0-3$; 
so does the group generated 
by the operators $K^{(\mu)}$. 
One can choose $K^{(2)}$  and $K^{(3)}$ 
as members of a complete set of commuting operators. 
Since $\left(K^{(2)}\right)^q=\left(K^{(3)}\right)^{q'}={\mathbf I}$ 
the Hilbert space can be 
decomposed into subspaces labeled by integers $1\le n\le q$,
 $1\le n'\le q'$.

Selecting $-\imath D^{(0)}_{1}/F$ and
$-\imath D^{(0)}_{0}/F'$  as position operators, we
express our fields as functions 
of two real variables, $h_{n,n'}(z,z')$. The indices $n$,$n'$ 
label the appropriate subspace.
The covariant derivatives act as follows:
\begin{eqnarray}
D^{(0)}_{1}h_n(z,z')&=&\imath F z h_n(z,z')\\
D^{(0)}_{0}h_n(z,z')&=&\imath F' z' h_n(z,z')\\
D^{(0)}_{2}h_n(z,z')&=&\partial_z h_n(z,z')\\
D^{(0)}_{3}h_n(z,z')&=&\partial_{z'} h_n(z,z')
\end{eqnarray}
The expression relating the initial four dimensional fields
$\psi(x)$ and $h_{n,n'}(z,z')$ can be obtained from (\ref{changebasis}).

\section{Uniform abelian fields} 
We can use less dimensions for the fields, but, 
except for special backgrounds, the
interactions will be non-local. If we are only 
concerned with fluctuations around
one of the abelian backgrounds which played a role earlier,
not only do the equations stay local, but, in addition,
the Dirac operator becomes evidently exactly diagonalizable.
As a result, the famous relations between topology
and fermionic zero modes is made very explicit.   

We begin in two dimensions with a gauge potential given by:
\be
\label{gauge2}
A_i(x)=-\pi q\epsilon_{i j}(x_j-2 x_j^{(0)})/{\cal A} 
\ee
The corresponding field is $F=F_{12}=2\pi q/(l_1 l_2)$.
The coordinates $x_i^{(0)}$  in Eq.~(\ref{gauge2})
specify the location where Polyakov loops 
around the torus take value $1$ and represent 
non-trivial parameters of the gauge background.
For $x_i^{(0)}=0$, the covariant derivative coincides with 
$D_i^{(0)}$ of section 2.
In general,
\be
D_i\equiv\partial_i-\imath A_i(x)={\cal T}_1(-x_1^{(0)}) {\cal T}_2(-x_2^{(0)})
D_i^{(0)}  {\cal T}_2(x_2^{(0)}) {\cal T}_1(x_1^{(0)})
\ee
Thus, the eigenvectors can always be obtained from
those at $x_i^{(0)}=0$.

The massless 
Dirac operator $D=\tau_1 D_1^{(0)} + \tau_2 D_2^{(0)}$, 
should have $q$ zero modes.  
Let $\tau_i$ denote the Pauli matrices used to express the 
Dirac matrices $\gamma_i$ in the chiral basis. 
The eigenvalue equation
\be
\left(\begin{array}{cc} 0 &   D_1^{(0)}-\imath  D_2^{(0)}\\
 D_1^{(0)}+\imath  D_2^{(0)} & 0 \end{array}\right) \ \left( \begin{array}{c}
\psi^{(\lambda)}_1 \\ \psi^{(\lambda)}_2 \end{array} \right) = 
\imath \lambda \ \left( \begin{array}{c}
\psi^{(\lambda)}_1 \\ \psi^{(\lambda)}_2 \end{array} \right) \quad 
\ee 
in the one-dimensional 
{\em position space} basis of section 2 is built out of:
\be
 D_1^{(0)}\pm \imath  D_2^{(0)} \longrightarrow  \imath (\partial_z \pm Fz)=
\left\{
\begin{array}{c} 
\imath \sqrt{2 F} a \\ 
\imath \sqrt{2 F} a^\dagger 
\end{array} \right.
\ee   
$\partial_z$ stands for the derivative with respect to $z$ and 
$a$ is a standard harmonic oscillator 
annihilation operator:
\begin{eqnarray}
a&=&\frac{1}{2}\sqrt{\frac{l_1l_2}{\pi q}}\partial_z+ \sqrt{\frac{ \pi q
}{l_1l_2}}z \nonumber \\
\label{aadag}
a^{\dag}&=&-\frac{1}{2}\sqrt{\frac{l_1 l_2}{\pi q}}\partial_z+
\sqrt{\frac{ \pi q}{l_1l_2}}z
\end{eqnarray}

Dirac's eigenvalue equation takes the form
\begin{equation}
\label{Dirac4}
\imath \sqrt{2 F}
\pmatrix{ 0 &   a^{\dag} \cr  a & 0} \pmatrix{h^{(\lambda)}_{1,n} \cr
h^{(\lambda)}_{2,n} } = \imath \lambda  \pmatrix{h^{(\lambda)}_{1,n} \cr
h^{(\lambda)}_{2,n} }
\end{equation}
with $h^{(\lambda)}_{i,n}$ representing the eigenmode wave function 
in the one dimensional basis. The index $n$ takes $q$
values and refers
to the internal degree of freedom representing eigenstates of $K^{(2)}$ showing
that  spectrum is $q$-fold degenerate. 

We express the expected positive
chirality zero-mode in each sector
in terms of $\vert p \rangle$, 
the eigenstate of the number operator  
$a^{\dag}a$:
\be
\pmatrix{\vert 0 \rangle \cr 0}
\ee 
The rest of the eigenstates are given by:
\begin{equation}
\label{spectrum}
h^{(\pm p)}_n = \frac{1}{\sqrt{2}}\pmatrix{\vert p \rangle
\cr \pm \vert p-1 \rangle}
\end{equation}
where $p=1,2 \dots$.   
The corresponding eigenvalues are
\begin{equation}
\label{eigen}
\imath \lambda^{(\pm p)} = \pm \imath \sqrt{2Fp}
\end{equation}
$h^{(\pm p)}$ map into each other 
under chirality: $\tau_3
h^{(\pm p)}=h^{(\mp p)}$. 

In the original two-dimensional 
representation we obtain a well known form of the zero-mode. Starting from
\begin{equation}
\label{zero2}
h^{(0)}_{1, n}(z) = \left(\frac{2q}{l_1 l_2}\right)^{1/4} e^{-\frac{F
z^2}{2}}
\end{equation}
we get, with
$(x,y)\in {\bf T^2}$,
\begin{eqnarray}
\label{zero3}
\psi^{(0)}_1(x_1,x_2) = 
\left(\frac{2q}{l_1 l_2}\right)^{1/4} \times &\nonumber \\
\exp\{\imath \pi q\frac{x_1 x_2}{{\cal A}}\}\ 
\sum_{n=1}^q\sum_{s\in{\mathbf Z}}\ & c_n\
e^{-\frac{F (x_2+\frac{n+qs}{q}l_2)^2}{2}+\imath F x_1 l_2 (n+sq)}.
\end{eqnarray}
The $c_n$ are the components of an arbitrary $q$-dimensional 
vector of unit modulus.

The zero-mode is not spatially homogeneous in terms of either 
$(x_1,x_2)$ or $z$.  In particular, for $q=1$, the solution is proportional 
to a theta function. 
It vanishes at the point $\bar{x}_1=l_1/2,\bar{x}_2=l_2/2 $: 
Eq.~(\ref{zero3}) gives
\begin{equation}
\label{midpoint}
\psi^{(0)}_1(\bar{x}_1, \bar{x}_2) = e^{\imath \pi/4} 
\left(\frac{2q}{l_1 l_2}\right)^{1/4}
  \sum_s
e^{-\frac{F l_2^2}{2}(s +\frac{1}{2})^2} (-1)^s
\end{equation}
The exponential is even under $s \to -1-s$ while
$(-1)^s$ is odd.  The inhomogeneity of the
zero-mode reflects the existence of the two extra physical
parameters $(x^{(0)},y^{(0)})$,  
beyond the magnitude of the quantized magnetic flux,
identifying where the Polyakov loops winding around the torus take value $1$. 
At the coordinates of the zero of $\psi^{(0)}_1$,
the two Polyakov loops take value -1. These coordinates could be used
as parameters as well.

According to Eqs.~(\ref{spectrum}),~(\ref{eigen})
the spectrum for $q>1$ is the same 
as the spectrum of $q$
independent fermions, each defined over a region of area
${\cal A}/q$.  This can be understood as follows:
Dirac's equation for $q>1$ could be solved by using the $q=1$
solution over the region $0 \le x_1 \le l_1/q$, $0 \le x_2 \le
l_2$ (which carries one unit of magnetic flux) and extending
it via the boundary condition $\psi(x_1+l_1/q,x_2)=\exp[\imath  \pi
x_2/l_2] \psi(x_1,x_2)$ to the rest of the $x_1$ domain $l_1/q \le x_1 \le
l_1$. This gives the solution corresponding to the sector $n=0$, 
an eigenspace of $K^{(2)}$. 
Applying $K^{(1)}$ one obtains the remaining $q-1$ other solutions.
All of them will satisfy the Dirac equation and the boundary condition
\begin{equation}
\label{qboundary}
\psi(x_1+l_1/q,x_2)=e^{\imath  \pi  x_2/l_2+\imath 2\pi n/q } \psi(x_1,x_2)
\end{equation}
Thus, the spectrum is that of $q$ independent
fermions defined over an area with a single unit of magnetic flux.

Let us turn to the  massive Dirac operator  $D+m$.
For $q=1$ we get:
\begin{equation}
\label{massive_dirac}
\pmatrix{ m & \imath \sqrt{2F} a^{\dag} \cr  \imath \sqrt{2F} a & m} \pmatrix{\tilde h^{(\lambda)}_{1} \cr
\tilde h^{(\lambda)}_{2} } = \lambda \pmatrix{\tilde
h^{(\lambda)}_{1} \cr \tilde h^{(\lambda)}_{2} }
\end{equation}

Since in the presence of mass $D+m$ is neither hermitian nor 
antihermitian it is better to consider the Hermitian Dirac
operator $H=\tau_3 D$:
\begin{equation}
\label{massive_dirac_ham}
\pmatrix{ m &  \imath \sqrt{2F} a^{\dag} \cr
- \imath \sqrt{2F} a & -m} \pmatrix{\tilde h^{(E)}_{1} \cr
\tilde h^{(E)}_{2} } = E \pmatrix{\tilde
h^{(E)}_{1} \cr \tilde h^{(E)}_{2} }
\end{equation}
$H$ has one isolated eigenvalue at $E=m$ with $m$-independent eigenfunction
$\pmatrix{ \vert 0 \rangle \cr 0}$ and an infinite set of
paired eigenvalues, each pair labeled by $p=1,2,3,\dots$.
\begin{eqnarray}
\label{spectrum_ham}
E^{(\pm p)} =& \pm\sqrt{m^2+2pF},~~~~
\tilde h^{(\pm p)} = \frac{1}{\sqrt{1+(z^{(\pm p)})^2}}
\pmatrix{\vert p \rangle
\cr z^{(\pm n)} \vert p-1 \rangle}\\
&z^{(+p)}z^{(-p)}=-1
\end{eqnarray}
with
\begin{equation}
z^{(\pm p)}=-\frac{m}{2} \sqrt{ \frac{l_1 l_2}{p\pi} } \pm
\sqrt{ \frac{m^2}{4} \frac{l_1 l_2 }{p\pi}+1}
\end{equation}

As in all other cases, the states and eigenvalues depend only
on the area ${\cal A}=l_1 l_2$ but not on the ratio $l_1/l_2$. One can
view the one dimensional version as the limit when the ratio
goes to zero or infinity while the area is kept fixed. 

When we vary $m$ from positive values to negative ones 
isolated state crosses zero energy at $m=0$.
The paired eigenvalues never get to zero and are symmetric
under $m\to -m$. The paired eigenstates depend on $m$, but the
signs of $z^{(\pm p)}$ are $m$-independent. On the lattice
there is full dependence on $m$, but the number of zero
level crossings as $m$ is varied stays the same as in continuum.

Consider now four dimensions with gauge group $SU(2)$ and
fields in the fundamental, defined over an $l_0\times l_1 \times l_2 \times l_3$ 
torus. The boundary conditions are taken in the canonical form from the beginning:
\be
Q^{\pm}_{\mu \nu}= \pm \pmatrix{0 & 0 & 0 & q'\cr
0 & 0 & q &0 \cr 0& -q & 0 &0\cr -q'& 0 & 0 & 0}   
\ee
$\pm$ denotes the two color components. Note that the 
first Chern class invariants vanish. The topological charge is  
given by $Q=2 q q'$ with the factor $2$ arises coming from the sum
over color components. Odd topological charges cannot be obtained 
this way - as mentioned before, $SU(2)$ is somewhat exceptional. 
The following constant field strength is 
compatible with the boundary conditions:
\be
\label{conFS}
F^{\pm}_{\mu \nu}= \pm \pmatrix{0 & 0 & 0 & F'\cr
0 & 0 & F &0 \cr 0& -F & 0 &0\cr -F'& 0 & 0 & 0}   
\ee
where $F=2 \pi q/(l_1 l_2)$ and $F'=2 \pi q'/(l_0 l_3)$.
The upper color component coincides with the matrix 
$F^{(0)}_{\mu \nu}$ of the previous section in 
canonical form.

The fermion field is in the fundamental representation of $SU(2)$ denoted 
by:
\be
\Psi=\left(\begin{array}{c} \psi ^{+} \\  \psi^- \end{array}\right)
\ee
The index theorem tells us that there are $Q=2 q q'$
zero-modes of positive chirality more than zero modes of negative 
chirality. The eigenvalue equations decouple into color components.
Because of the pseudoreality of $SU(2)$ the upper color component
determines the lower one by the symmetry
\be
\Psi \longrightarrow \tau_2^S \tau_2^C \Psi^*  
\ee
$\tau_2^{S,C}$ act in spin 
and color space respectively while $\Psi^*$ is the complex 
conjugate vector. The spectrum is doubly degenerate. 
We choose a chiral representation for the Dirac matrices:  
\be
\gamma_{\mu}=\left( \begin{array}{c c} 0 & \sigma_{\mu} \\
\overline{\sigma}_{\mu} & 0 \end{array} \right)
\ee
\be
 \gamma_5=\left( \begin{array}{c c} {\mathbf I} & 0 \\
0 & -{\mathbf I} \end{array} \right) 
\ee
The matrices $\sigma_{\mu}=({\mathbf I},-\imath\vec{\tau})$ are the Weyl
matrices and $\overline{\sigma}_{\mu}$ are the adjoints. 
They satisfy:
\be
\overline{\sigma}_{\mu} \sigma_{\nu} = \bar{\eta}_{\mu \nu}^{\alpha}
\sigma_{\alpha}\quad 
\ee
$\bar{\eta}_{\mu \nu}^{\alpha}$ is the `t Hooft symbol; 
$\overline{\eta}^{0}_{\mu \nu}=
\delta_{\mu \nu}$ and the  $\overline{\eta}^{i}_{\mu \nu}$ are a basis of 
antiself-dual tensors. The eigenvalue problem is:
\be
D  \psi(x) \equiv D_{\mu}\gamma_{\mu} \psi(x) = \pmatrix{0&\overline{D}\cr\widehat{D} &0 }
\psi(x)= \imath \lambda \psi(x) 
\ee
The symbol $\overline{D}=D_{\mu} \overline{\sigma}_{\mu}$
($\widehat{D}=D_{\mu} \sigma_{\mu}$ ) denotes the 
positive (negative) chirality Weyl operator restricted to 
the upper color component. The covariant derivative $D_{\mu}$
coincides with the generator of translations studied in the last 
section, $D_{\mu}^{(0)}$. 
Choosing a new origin of coordinates gives the vector potential
another form. The 
different choices are gauge inequivalent, as evidenced by the Polyakov loops, 
but are related by translations ${\cal T}(b)$.

In position space representation $\overline{D}$
becomes ($\widehat{D}=-\overline{D}^\dagger$):
\be
\label{Dbar}
\overline{D}=\pmatrix{D_0+\imath D_3 & \imath(D_1-\imath D_2)\cr
\imath(D_1+\imath D_2) & D_0-\imath D_3}\longrightarrow 
\pmatrix{\imath \sqrt{2F'} a' & -\sqrt{2F} a^\dag \cr 
-\sqrt{2F} a & \imath \sqrt{2F'} a'^\dag}
 \ee
$a$,$a'$ are two independent annihilation operators.
$a$ is the same as in Eq.~(\ref{aadag}). 
To get $a'$ replace the $1-2$ plane 
by the $0-3$ plane and $z$ by $z'$.

A positive chirality eigenstate of the Dirac operator with eigenvalue $\imath 
\lambda$, also is an eigenstate of the operator $\widehat{D}\overline{D}$ with 
eigenvalue $-\lambda^2$:
\be
\widehat{D}\overline{D}=-2\pmatrix{F'a'^\dag a'+F a^\dag a & 0 \cr
0 & F' a' a'^\dag +F a a^\dag}
\ee
Let $\vert p,p'\rangle$, with non-negative integers $p$ and $p'$,
be simultaneous eigenvectors
of the number operators. 
There is a  positive chirality zero-mode given by: 
\be
h_{n,n'}^{(0)}=\pmatrix{\vert 0 , 0 \rangle \cr 0 \cr 0 \cr 0} 
\ee
There is an identical zero-mode in each of the subspaces labeled
by $n$-$n'$ making up $q q'$ zero-modes for each color 
component and confirming the index theorem. 
Actually, all operators act in  
the same way in each subspace, so the entire spectrum is replicated $Q$ times.

The set of eigenvalues of the Dirac operator consists of:
\be
\imath\lambda^{(\pm,p,p')}= \pm \imath \sqrt{2Fp+2F'p'} 
\ee   
The corresponding eigenvectors are:
\be
\frac{1}{\sqrt{2}} \pmatrix{ \vert p , p' \rangle \cr
0 \cr \pm \sqrt{\frac{F'p'}{Fp+F'p'}}\, \vert p , p'-1 \rangle \cr 
\pm\imath \sqrt{\frac{Fp}{Fp+F'p'}}\, \vert p-1 , p' \rangle }
\ee 
For $p,p'>0$ there exist additional eigenstates with the same eigenvalue:
\be
\frac{1}{\sqrt{2}}  \pmatrix{ 0 \cr \vert p-1 , p'-1 \rangle \cr
 \pm\imath  \sqrt{\frac{Fp}{Fp+F'p'}}\, \vert p , p'-1 \rangle \cr
\pm \sqrt{\frac{F'p'}{Fp+F'p'}}\, \vert p-1 , p' \rangle }
\ee  

The spectrum and eigenstates of  the massive hermitian Dirac operator
 $H=\gamma_5(D+m)$
can be easily obtained in terms of those of 
the massless Dirac operator. The unchanged zero-modes become
eigenstates of $H$ with  eigenvalue $m$. The massive Dirac operator
mixes other eigenstates 
of the massless Dirac operator, $\psi_\lambda$, with $\gamma_5 \psi_\lambda$. 
The eigenvalues and eigenvectors of $H$ result  from diagonalizing 
the corresponding $2\times 2$ matrix giving 
eigenvalues equal to $\pm\sqrt{m^2+\lambda^2}$. 
The eigenvectors are of the form 
$\frac{1}{\sqrt{2}}(e^{-\imath \delta/2}\psi_\lambda \pm
e^{\imath \delta/2}\gamma_5\psi_\lambda)$, where $\delta $ is the phase 
of the complex number $m +\imath \lambda$.

\section{Two dimensions -- lattice}
\label{discrete}
We now turn to the lattice and examine a constant abelian background in
two dimensions. We wish to determine how an ultra-violet cutoff interacts with
the ``dimensional reduction'' we saw before.  

The fermion fields $\psi(n_1,n_2)$ 
are complex functions of sites
on an $L_1\times L_2$  square  lattice. 
If we visualize the computer memory storing
these functions, there is only a finite list of complex numbers 
and no mention of boundary conditions. 
No continuum features are lost
because the link gauge variables are parallel
transporters to a finite distance and therefore in the continuum 
limit include both the vector potential 
and the transition functions. The link variables
thus keep track also of what we normally view as boundary conditions.
To keep our lattice description formally close 
to the continuum formulation we choose to view our lattice fields
slightly differently,  
as defined over an infinite lattice but restricted by:
\be
\label{LatBCTD}
\psi(n+E^{(i)})= \exp\{\imath \pi \sum_{j} q_{i j }
\frac{n_{j}}{L_{j}}\}\; \psi(n)
\ee 
$E^{(i)}$ is a vector of length $L_{i}$ in the $i$ direction, 
$q_{i j}=q\epsilon_{i j}$ and the indices $i,j$ take the values $1,2$. 

The 
space of fields satisfying the boundary conditions~(\ref{LatBCTD}) can
be described in a way 
equivalent to the continuum one-dimensional quantum mechanical 
formulation.     
First, we gauge-transform  the function $\psi(n)$:        
\be
\kappa(n)=\exp \{  -\pi \imath q \frac{n_1 n_2}{L_1 L_2} \}\, \psi(n)
\ee 
The new function, $\kappa$, is periodic in $n_1$ with period $L_1$
and can be parametrised as:
\be
\kappa(n_1,n_2)=\sum_{p_1=1}^{L_1}\, \exp\{\imath 2 \pi p_1 n_1/L_1\}\,
c(n_2,p_1)
\ee 
The momentum $p_1$ is defined modulo $L_1$. 
Imposing the boundary condition in  $n_2$,  we get:
\be
c(n_2+L_2,p_1)=c(n_2, p_1+q)
\ee 
Defining  $q_1=\gcd(q,L_1)$, we can write $p_1=sq+n \bmod L_1$ where $n$ is
an integer defined modulo $q_1$
and $s$ is an integer defined modulo $L_1/q_1$:
\be
c(n_2,p_1)=h(n_2+L_2s, n)
\ee
We have arrived at the following expression for $\psi(n)$:
\begin{eqnarray}
&\psi(n_1,n_2)= 
\exp \{  \pi \imath q \frac{n_1 n_2}{L_1 L_2} \} \times 
\nonumber \\
&\sum_{s=1}^{L_1/q_1}\sum_{n=1}^{q_1} 
\exp\{\imath 2 \pi (sq+n)
n_1/L_1\}\, 
h(n_2+L_2s, n)
\end{eqnarray}
For q=1, $q_1=1$, and our representation is given
in terms of a function $h$ defined on a one-dimensional lattice 
of length $N\equiv L_1 L_2$. For general $q$, it is still 
true that the space of fermions  fields is $N$-dimensional, but now
one has  $q_1$ functions living on a periodic lattice of size 
$N/q_1$.

The lattice fields $\psi(n)$ can be viewed as the collection of 
values of the continuum fields $\psi(x)$ (appearing in Section 2) at 
the lattice points $x=na$, where $a$ is the lattice spacing and 
$l_{\mu}= L_{\mu} a $. Thus, the transformations acting on the 
space of lattice fields   are  given by the restriction of the 
continuum transformations  to the lattice points. In particular, we have a 
group of modified translations by vectors whose components are
integer multiples of the lattice spacing $a$. The generators of the 
group are ${\cal T}_{i}$: 
\be
{\cal T}_{i}\psi(n)=\exp\{\imath \pi \sum_j \frac{q_{i j } n_j}{N}\}\
\psi(n+\hat{\imath})
\ee  
$N$ equals $L_1L_2$ and $\hat{\imath}$ is a vector of unit length in the 
$i$ direction. We obtained a non-commutative discrete subgroup of the continuum 
translation group:
\be
{\cal T}_1{\cal T}_2=\exp\{-2 \pi
\imath \frac{q}{N}\}\  {\cal T}_2{\cal T}_1
\ee
The one-dimensional representation is obtained in a basis that diagonalizes 
${\cal T}_1$:
\be
{\cal T}_1 h(k,n)=\exp\{ 2 \pi \imath (\frac{q k}{N} + \frac{n}{L_1})\}\ h(k,n)
\ee
The translation operator in the $2$ direction acts as follows:
\be
{\cal T}_2 h(k,n)= h(k+1,n) 
\ee
  
If $N_{1 2}\equiv\gcd({N,q})\ne 1$ the group
generated by ${\cal T}_{i}$ acts  reducibly  
on our space of lattice fields decomposing it  
into $N_{1 2} $ subspaces, each invariant under translations. 
This constitutes the lattice 
remnant of the  group  generated by $K^{(i)}$ in the continuum. 
These operators contain translations by $l_i/q$ and 
can be extended  to the lattice 
only if $L_i$ is divisible by $q$. Otherwise, one obtains only a subgroup 
of the group found in the continuum.  

We pick lattice gauge fields having a constant plaquette
value:
\be
U_{i}(n)=\alpha_{i}\, \exp\{\imath \pi \sum_{j} \frac{q_{i j} n_{j} }{N}\}
\ee
The $\alpha_{i}$ are complex numbers of unit modulus. 
The plaquettes are given by:
\be
U_{1 2}(n)= \exp\{-\imath 2 \pi  \frac{q}{N}\}\equiv \zeta^{-q}
\ee
where $\zeta=\exp\{\imath 2 \pi /N\}$. The values of the
plaquettes do not depend on $\alpha_{i}$, but the latter 
are physical parameters because they influence 
the Polyakov loops. The $\alpha_i$ 
are related to the continuum parameters $x_i^{(0)}$ (taken in units of
the lattice spacing):
\be
\alpha_{i}=e^{ -2 \pi \imath q \sum_j \epsilon_{i j} x_j^{(0)}/ N}
\ee 
In the continuum the spectrum of the Dirac operators was independent 
of $x_i^{(0)}$, by unitary equivalence generated by 
appropriate translations. On the lattice
translations are  discrete and can only be used
to restrict the ranges of the $x_i^{(0)}$:  $0 \le  x_i^{(0)}<  1$.

The lattice Dirac operators contain the covariant shift operators $T_i$:
\be
T_i=\alpha_i {\cal T}_i 
\ee 

For the rest of the section we restrict ourselves to the $q=1$ case.
The $T_i$ operators take the
following matrix form when acting on $h(k)$'s:
\begin{equation}
\label{Xmatrix}
T_2= \alpha_2 \pmatrix{0 & 1 & 0 & \dots & 0 \cr
                  0 & 0 & 1 & \dots & 0 \cr
                  \dots  & \dots &\dots &\dots & \dots   \cr
                  0 & 0 & \dots & 0 & 1 \cr
                  1 & 0 & \dots & 0 & 0  }
\end{equation}
\begin{equation}
\label{Ymatrix}
T_1= \alpha_1 \pmatrix{1 & 0 & 0 & \dots & 0 \cr
                   0 & \zeta & 0 & \dots & 0 \cr
                   0 & 0 &\zeta^2 & \dots & 0 \cr
                  \dots  & \dots &\dots &\dots & \dots   \cr
                   0 & 0 & \dots & 0 & \zeta^{N-1}  }
\end{equation}

It is convenient to choose a different basis of Dirac matrices for the lattice
analysis:
\begin{equation}
\label{lattice-gamma}
\gamma_1=\tau_2, \gamma_2=\tau_3, \gamma_3=-i\gamma_1\gamma_2=\tau_1
\end{equation}
The Wilson Dirac operator $D_W$ is
\begin{equation}
\label{d-w}
D_W=m+\sum_\mu (r-V_\mu)
\end{equation}
where the unitary matrices $V_\mu$ are:
\begin{equation}
\label{v-mu}
V_\mu={{r-\gamma_\mu}\over 2} T_\mu +{{r+\gamma_\mu}\over 2} T_\mu^\dagger
\end{equation}
In what follows we will fix the Wilson parameter $r$ to $1$.
{}From now on periodicity in the ``site'' index $n$ is implied
by the convention that it is always viewed modulo $N$ and this
extends to operations involving site indices. For simplicity, let
us start with $x_i^{(0)}=0$.

The Wilson Dirac lattice Hamiltonian is
\begin{equation}
\label{h-w}
H_W=\tau_1 D_W= \pmatrix{-\sin\Theta& 2+m -\cos\Theta - T\cr
      2+m-\cos\Theta  - T^\dagger & \sin\Theta  \cr}
\end{equation}
Here, $\Theta$ is an $N\times N$ diagonal matrix:
\begin{equation}
\Theta=diag[1,\theta_1,\theta_2,\dots,\theta_{N-1}]
\end{equation}
with
\begin{equation}
\theta_n={{2\pi n}\over N},~~ T_{kn}=\delta_{k+1,n}
\end{equation}

$H_W$ is a local one dimensional Hamiltonian defined on a 
discretised circle. The calculation
of eigenvalues and eigenvectors requires fewer operations
than in the basis corresponding to a two dimensional grid.

Writing the eigenvector for energy $E$ as 
\begin{equation}
\label{psi}
\psi_n = \pmatrix{a_n\cr b_n}
\end{equation}
where $n$ ranges between $0$ and $N-1$, we get the eigenvalue equations
\begin{eqnarray}
\label{recursion}
&Ea_n=-\sin\theta_n a_n + (2+m-\cos\theta_n )b_n - b_{n+1}\\
&Eb_{n+1} =(2+m-\cos\theta_{n+1} )a_{n+1} -a_n +\sin\theta_{n+1} b_{n+1}
\end{eqnarray}
In matrix form we have
\begin{eqnarray}
\label{recurs-matrix}
\pmatrix{2+m-\cos\theta_{n+1}&\sin\theta_{n+1} -E\cr 0&1\cr}
\pmatrix{a_{n+1}\cr b_{n+1}\cr} =\\
\nonumber
\pmatrix{1&0\cr -E-\sin\theta_n & 2+m -\cos\theta_n\cr }
\pmatrix{a_n\cr b_n}
\end{eqnarray}
Inverting the matrix on the left hand side we get
\begin{equation}
\pmatrix{a_{n+1}\cr b_{n+1}\cr} =M_n \pmatrix{a_n\cr b_n}
\end{equation}
The eigenvalue condition is that $M\equiv M_{N-1}M_{N-2}\dots M_0$
have an eigenvalue equal to unity. Observe that $\det M =1$; 
thus if $M$ has one eigenvalue unity the other
also must be unity and the eigenvalue condition
can also be written as
\begin{equation}
Tr M(E)=2
\end{equation}
When this condition is fulfilled $M(E)$ has a single eigenvector,
not two, because it is non-diagonalizable. The energies $E$
are the roots of a polynomial of degree $2N$. Typically these roots are
all simple and to each corresponds a single eigenstate of $H_W$ --
there are no degeneracies.  

We have ended up with an exact formula for the characteristic polynomial of $H_W$. 
\begin{equation}
\label{char-poly}
\det(E-H_W)=(-1)^{N} K_N [Tr M(E) -2]
\end{equation}
The proportionality factor $K_N$ 
is calculated by comparing the terms
$E^{2N}$ in $Tr M(E)$ and in $\det (E-H_W)$.
\begin{eqnarray}
\label{k-n}
K_N =\prod_{n=0}^{N-1} (2+m-\cos \theta_n ) =&\\
\nonumber
\left [ {{2+m+\sqrt{(2+m)^2 -1}}\over 2}\right ]^N&
+\left [ {{2+m-\sqrt{(2+m)^2 -1}}\over 2}\right ]^N -{1\over {2^{N-1}}}
\end{eqnarray}
This formula   is correct  even when the square roots are imaginary, 
with the understanding that the second term on the right hand side 
is the complex conjugate of the first. It is derived by setting $2+m=\frac{1}{2}
(z+\frac{1}{z})$ and noting that the finite Laurent series in $z$ defining $K_N$
is symmetric under $z \leftrightarrow \frac{1}{z}$ and vanishes when $z$ is an
$N^{\rm th}$ root of unity. 

We can now easily re-introduce arbitrary $x_1^{(0)}$ and $x_2^{(0)}$.
All that happens is that $\Theta$ gets shifted to $\Theta-\frac{2 \pi
  x_1^{(0)}}{N}$
and the characteristic polynomial is now proportional
to
$$
Tr M(E)-2\cos (2\pi x_2^{(0)}) 
$$
The proportionality constant now depends on $x_1^{(0)}$.

Numerically, the finite precision of the machine will limit
the usefulness of a direct implementation of the characteristic
polynomial as a means of locating the eigenvalues of $H_W$ when
$N$ gets large. A better way exists for $x_i^{(0)}=0$ 
which has extra symmetry. $T_1$ and $T_2$ are
then isospectral and there exists an intertwining map, $F_N$.
(Actually, $T_1 , T_1^\dagger$, $T_2 , T_2^\dagger$ all have
identical spectra.)
$F_N$ is the discrete Fourier transform:
\begin{equation}
F_{jk}={1\over \sqrt{N}} e^{{{2\pi i}\over N}jk}
\end{equation}
$F$ is symmetric and unitary.
\begin{eqnarray}
T_2 F= FT_1\\ F T_2 =  T_1^\dagger F
\end{eqnarray}
$F$ can be combined with a rotation in spinor space to
produce a unitary matrix that commutes with $H_W$.
\begin{equation}
U={{1-i}\over\sqrt{2}} {{1-i\gamma_3}\over \sqrt{2}} F
\end{equation}
One checks now that
\begin{eqnarray}
D_W U= U D_W\\
U H_W=H_W U 
\end{eqnarray}
The simpler case of commutativity with 
$U^2$ is more obvious and can be checked
directly:
\begin{eqnarray}
F^2 e^{i\Theta} F^2 =e^{-i\Theta}
\end{eqnarray}
leading to
\begin{eqnarray}
F^2 H_W F^2 =\tau_1 H_W \tau_1
\end{eqnarray}
where we notationally ignore factors of unit matrices in tensor
products.

The eigenvalues of $U$ are $1,i,-1,-i$.
Their degeneracies are computed from the ranks of the appropriate 
projectors. Since $U^4=1$ the projectors can be written in terms
of $1,U,U^*,U^2$. For the multiplicities we only need the traces
of $1,F,F^*, F^2$. The trace of $F^2$ is trivial because
\begin{equation}
F^2=\pmatrix{1&0&0&0&\dots&0&0\cr
             0&0&0&0&\dots&0&1\cr
             0&0&0&0&\dots&1&0\cr 
             \vdots&\vdots&\vdots&\vdots&\dots&\vdots&\vdots\cr
             0&0&1&0&\dots&0&0\cr
             0&1&0&0&\dots&0&0\cr}
\end{equation}
The trace of $F$ is not trivial and is given by a famous sum of Gauss:
\begin{equation}
S_G(N)={1\over\sqrt{N}} \sum_{k=0}^{N-1} e^{
\frac{2\pi \imath }{N} k^2}
\end{equation}
$S_G(N)$ is determined by its values at $N=1,2,3,4$ and by
(nontrivial) mod 4 periodicity: $S_G (1)=1,~S_G(2)=0,~S_G(3)=i,~S_G(4)=1+i$. 
The reflection $U^2=\gamma_3 F^2=\tau_1 F^2$
reverses the order of sites with $n=0$ as a fixed point
for all $N$ and flips $a$ and $b$ (in our $\gamma$-matrix basis).

The origin of the $U$-symmetry is in the continuum.  
The harmonic oscillator case is unique in that
all the eigenfunctions are eigenstates of the Fourier transform
as a result of a discrete symmetry that exchanges the coordinate with
its conjugate momentum.
The eigenvalues of the Fourier transform 
are $1,i,-1,-i$ depending on the oscillator level
mod 4. To get an operator that commutes with the Dirac
operator in our case 
the Fourier transform needs to be combined
with the action of a rotation in spinor space depending on 
the chirality matrix $\gamma_3$. On the lattice one can 
decompose $H_W$ into four
blocks each of sizes (roughly) of $N/2\times N/2$. The exact sizes
depend on $N$ mod 4. The $U$-symmetry is simpler to understand when
also using the standard representation in terms of fields on a two
dimensional torus of equal sides. Then we have a symmetry under rotations by
$\frac{\pi}{2}$, which corresponds to a switch between coordinate and momentum
in the one-dimensional framework. 
The two dimensional interpretation as a discrete rotation
explains why an action on spinorial indices is also required. 
As only the area is relevant, the requirement that
the torus have equal sides can be dropped. In practice we shall exploit only
$U^2$, corresponding to a rotation by $\pi$, and the issue does not even arise. 
In the one dimensional basis the $U^2$ 
reflection symmetry contains the
square of the Fourier transform combined with $\gamma_3$ in
a simple manner. 

The $U^2$ symmetry allows us to factorize $Tr M(E) -2$. The trace
contains the sum of the expectation of $M(E)$ in two states. These
contributions are individually high degree polynomials in $E$ and vary over many orders
of magnitude. The contributions nearly cancel at the eigenvalues and are
very sensitive to $E$. The symmetry makes it possible to work in only
one sector, where a single state contributes. As a result one deals with
a single expectation value, which can be renormalized iteratively, in effect
dividing out the quantity that has to vanish by a positive function of $E$ that
tames the large variability. This provides a numerically stable method
for locating the eigenvalues down to machine accuracy.

One needs to separate the case $N$=even
from $N$=odd. For $N=2L$, $L$-integer, the reflection has two fixed points:
$n=0$ and $n=L$. Thus, one can impose on $\psi_0$ and $\psi_L$
to be eigenvectors of $\tau_1$ with identical
eigenvalue. Start from $\psi_0={1\over\sqrt{2}}
\pmatrix {1\cr 1}\equiv \xi_+$ and let $\xi_-={1\over\sqrt{2}}
\pmatrix {1\cr -1}$. We need to have then
\begin{equation}
\xi_-^\dagger M_{L-1} (E) M_{L-2} (E) \dots M_0 (E)\xi_+ =0
\end{equation}
This gives a polynomial equation for $E$ of degree $2L$; the roots
are all the energies of the even states. Similarly, for the odd states
we have the equation:
\begin{equation}
\xi_+^\dagger M_{L-1} (E) M_{L-2} (E) \dots M_0 (E)\xi_- =0
\end{equation}
Extending to $N=2L-1$ requires imposing 
$\psi_{L-1} =\pm\tau_1 \psi_L$ coordinated with $\pm\tau_1\psi_0=\psi_0$. 
This implies
\begin{equation}
(1-\tau_1 M_{L-1} )\psi_{L-1} = 0
\end{equation}
if $\tau_1 \psi_0 = \psi_0$, and
\begin{equation}
(1+\tau_1 M_{L-1} )\psi_{L-1} = 0
\end{equation}
if $\tau_1 \psi_0 =- \psi_0$.

Explicit evaluation of $M_{L-1}$ produces:
\begin{equation}
M_{L-1}=\pmatrix{ {{1-(s+E)^2}\over {2+m+c}} & s+E\cr
         -(s+E) & 2+m+c\cr}
\end{equation}
where $s=\sin{\pi\over{2L-1}}$ and $c=\cos{\pi\over{2L-1}}$. 
The eigenvalues of $\tau_1 M_{L-1}$ are $\lambda=\pm 1$, independently of $E$.
The corresponding two eigenvectors do depend on $E$:
\begin{equation}
\hat\xi_\lambda (E) = {1\over \sqrt{1+[(\lambda+E+s)/(2+m+c)]^2}}
\pmatrix{1\cr {{\lambda+E+s}\over {2+m+c}}}
\end{equation}
The eigenvalue conditions are that $(M_{L-2} M_{L-3}\dots M_0 )\xi_\sigma$
be parallel to $\hat\xi_\sigma (E)$ for $\sigma=\pm 1$.
\begin{eqnarray}
&\hat\xi^T_{+1} (E) \tau_2 (M_{L-2} M_{L-3}\dots M_0 )\xi_+=0\\
&\hat\xi^T_{-1} (E) \tau_2 (M_{L-2} M_{L-3}\dots M_0 )\xi_-=0
\end{eqnarray}
They look slightly different from the even case because now the
vectors $\hat\xi_{\pm 1}$ are not orthogonal to each other. 

The above is useful for numerical searches for energy eigenvalues
of $H_W$. The vectors obtained
by the sequential action of $M_n$ are normalized at each step and
this eliminates accuracy problems. $N$ is limited now only because for very
large $N$ the spacings between some eigenvalues may be below machine accuracy.  

Looking at some examples the following patterns emerge: 
The odd and even states have energy sequences that
separate (interlace with) each other. For even $N$ the 
topological charge $-\frac{1}{2}Tr H_W /\sqrt {H_W^2}$ is carried
by the odd states. For odd $N$ the topological charge is carried by
the odd and even states equally - each sector having an imbalance equal 
to unity. When $m$ is taken from positive values to $\sim -1$
it is always an even state that does the zero crossing. There are no degeneracies. 

Since there are no degeneracies one can calculate directly the
eigenvalues $\lambda$ 
of the symmetry operator $U$ for each one of the
eigenstates.  A simple way to do this is
to evaluate $F\psi$ at site $0$ only. We distinguish the even
and odd cases. 

If the state is even, $\gamma_3 F^2 \psi =\psi$, 
then
\begin{equation}
(F\psi)_0={1\over\sqrt{N}} \sum_{n=0}^{N-1} \psi_n = \lambda
\psi_0
\end{equation}
and $\lambda$ can be $\pm 1$. 
If the state is odd, $\gamma_3 F^2 \psi =-\psi$, then
\begin{equation}
(F\psi)_0={1\over\sqrt{N}} \sum_{n=0}^{N-1} \psi_n = i\lambda
\psi_0
\end{equation}
and $i\lambda$ can be $\pm 1$.

We now turn to the naive lattice Dirac operator.
In the one-dimensional version we now have:
\begin{equation}
H_N = \pmatrix {-\sin\Theta  & {1\over 2}(T^\dagger - T )\cr
{1\over 2}(T - T^\dagger ) & \sin\Theta \cr}
\end{equation}
If the $L_i$ are even there are two additional 
symmetries
involving staggering operators along each direction:
\begin{equation}
S_i \psi(n) = (-1)^{n_i}\psi(n)\quad. 
\end{equation}   
The $S_i$ commute among themselves, satisfy $S_i^2=1$ and also: 
\begin{equation}
S_i T_i = -T_i S_i\quad. 
\end{equation}  
The two anticommuting matrices $\Pi_i=S_i \gamma_i$, $i=1,2$ commute
with $H_N$. We diagonalize one of 
the $\Pi_i$'s, $\Pi_2$ for example. The eigenspaces of the two 
$\pm 1$ eigenvalues 
correspond to two species of staggered fermions.  
Going over to the one-dimensional description, 
the operator $S_2$ becomes  the corresponding one dimensional 
staggering operator $\bar{S}$:
\begin{equation}
\bar{S}_{nm}=(-1)^n \delta_{nm}
\end{equation}
$\bar{S}^2=1$, $\bar{S}$ is hermitian and 
$\bar{S}T\bar{S}=-T$. 

We obtain a reduced Hamiltonian for naive fermions
by introducing the 
following hermitian-unitary matrix: 
\begin{equation}
R=\frac{1+\tau_1}{2}I+\frac{1-\tau_1}{2}S_2  
\end{equation}  
It brings $H_N$ into a simpler form:
\begin{equation}
RH_NR=-\tau_3 H_{NR} \quad
\end{equation}
where $H_{NR}=-\imath S_2 D_{12}$ 
and 
\begin{equation}
\label{DOT_Def}
D_{1 2} = \frac{1}{2}(T_1-T_1^\dagger +\imath T_2 -\imath T_2^\dagger)  
\end{equation}
In the one dimensional representation we get:
\begin{equation}
H_{NR}= \bar{S} ({1 \over 2} (T-T^\dagger)+\sin\Theta)
\end{equation}
The operator $S_1 S_2$ anticonmutes with $H_{NR}$,
implying that the spectrum of $H_N$ is real, symmetric around 
zero and doubly degenerate. We can cast the 
eigenvalue equation in the following form:
\begin{equation}
\pmatrix{a_{n+1}\cr b_{n+1}\cr} =\bar{M}_n(E) \pmatrix{a_n\cr b_n}
\end{equation}   
where 
\begin{equation}
\bar{M}_n(E)=\pmatrix{2E(-1)^n-2 \sin\theta_n & 1\cr 1& 0\cr} 
\end{equation}   
For $E$ to be an eigenvalue, 
$\bar{M}(E)=\bar{M}_N(E)\cdots \bar{M}_1(E)$ must have an  eigenvalue equal to $1$. 
However, since the determinant of $\bar{M}(E)$ is  $1$, the condition 
becomes again:
\begin{equation}
Tr(\bar{M}(E))=2\quad. 
\end{equation} 
An analysis similar to the one performed for the Wilson-Dirac operator 
would apply also here.

Let us conclude this section by analyzing the existence and multiplicity 
of zero modes. We restrict our attention to positive
chirality modes. There always exists an equal  number 
of  negative chirality modes; this is a reflection of the famous phenomenon of 
fermion doubling. The equation for the 
zero mode is 
\begin{equation}
\label{iterNZM}
h^{(0)}_{n+1}-h^{(0)}_{n-1}=-2 h^{(0)}_n \sin\theta_n 
\end{equation}  
This is a second order recursion relation determining all
the elements of the series in terms of $h^{(0)}_0$ and $h^{(0)}_1$. 
Our solutions must satisfy the periodicity condition 
$h^{(0)}_{n+N}=h^{(0)}_{n}$. It is necessary and sufficient 
that this condition be
satisfied for two consecutive values of $n$.  
Since the problem is homogeneous this could lead to two 
conditions  on the ratio   $h^{(0)}_1/h^{(0)}_0$, so existence is 
not guaranteed.  To prove  existence we make use  of the fact that the
solution, if it exists,  satisfies  $h^{(0)}_{n}=h^{(0)}_{-n}$. For $n=1$
this relation follows from the vanishing of $\theta_0$. For other values 
of $n$ it can be proven by applying  the iteration Eq.~(\ref{iterNZM}).

Most of our previous formulas have been derived for  even $N$, so 
we will consider this case first.
Then, $h^{(0)}_{N/2}=h^{(0)}_{-N/2}$, which is equivalent to 
periodicity at $n=-N/2$. From the vanishing of $\sin\theta_{N/2}$ we conclude 
$h^{(0)}_{N/2-1}=h^{(0)}_{N/2+1}$, which together with reflection around 
$n=0$ yields $h^{(0)}_{-N/2+1}=h^{(0)}_{N/2+1}$ (periodicity at $-N/2+1$).
We conclude that the iteration procedure is automatically periodic 
for arbitrary values of  $h^{(0)}_0$ and $h^{(0)}_1$ and therefore there are 
two positive chirality and two negative chirality zero-modes. Their 
shape can be automatically determined by applying the iteration 
to any pair of initial values of $h^{(0)}_0$ and $h^{(0)}_1$. Once one 
zero-mode  has been determined the other can be found by applying the 
symmetry operator:
\begin{equation}
 h_n^{(0)} \longrightarrow (-1)^n\, h_{N/2+n}^{(0)}\quad 
\end{equation}  

In the odd  $N$ case, although staggering does not apply, the equation for  the zero-modes~(\ref{iterNZM}) 
and  the reflection property   ($h^{(0)}_{n}=h^{(0)}_{-n}$) still holds. This, however, 
is not enough to imply periodicity, but allows to reduce the two equations to one. 
This  fixes the ratio $h^{(0)}_1/h^{(0)}_0$: A single zero-mode exists for each chirality.
To conclude our analysis of zero-modes we mention that for the $x_i^{(0)}\ne 0$ case, the 
reflection property and the existence of zero-modes is lost.

In conclusion, the one dimensional character of the lattice problem
holds independently of which lattice fermions we use. Also, from
the numerical point of view there are no hidden features and the
problem truly has the complexity of a one dimensional system.

\section{Four dimensions -- lattice}

In this section we look at an $SU(2)$ gauge theory
on a 4 dimensional lattice ${\cal L}$ of size  
$L_0\times L_1\times L_2\times L_3$. 
We again take
abelian transition functions:
\be
\label{LfourBC}
\psi(n+E^{(\mu)})= \exp\{\imath \pi  \, \sum_{\nu} q_{\mu \nu} \frac{n_{\nu}}{L_{\nu}}
\tau_3\}\ \psi(n)
\ee 
$q_{\mu \nu}$ is an antisymmetric tensor whose only non-zero
components are $q_{0 3}=q'$ and $q_{1 2}=q$. 
The vector $E^{(\mu)}$  is a vector of length $L_{\mu}$ along the
$\mu$-direction.  Following the two dimensional example
we introduce shift-operators ${\cal
  T}_{\mu}$ which act as follows:
\be
{\cal T}_{\mu}\psi^+(n)=\exp\{\imath \pi  \, \sum_{\nu} q_{\mu \nu}  
\frac{n_{\nu}}{L_{\mu}
  L_{\nu}} \}\, \psi^+(n+\hat{\mu})
\ee 
Changing the sign of the  exponent  on the left-hand side, one obtains the
translation operator acting on the lower color component.  The
translation operators are just 
the covariant shift-operator in a background lattice gauge field with 
the links given by:
\be
\label{ConstGF}
U_{\mu}(n)=\exp\{\imath \pi  \,  \sum_{\nu} q_{\mu \nu} \frac{n_{\nu}}{L_{\mu} L_{\nu}}\tau_3\}
\ee
These links produce a homogeneous plaquette field. 

The covariant shift operators define a finite group:
\be
{\cal T}_{\mu} {\cal T}_{\nu} =\exp\{-2 \pi \imath \frac{q_{\mu \nu}}
{L_{\mu} L_{\nu}} \}\, {\cal T}_{\nu} {\cal T}_{\mu}
\ee 
For $q=q'=1$ the entire  $L_0L_1L_2L_3$-dimensional space 
is irreducible under this group. A convenient basis is 
obtained by diagonalising a maximal set of commuting matrices; 
we choose $ {\cal T}_{1}$ and $ {\cal T}_{0}$, leading to a two
dimensional  labeling of the basis given by the corresponding 
 eigenvalues of $ {\cal T}_{1}$ and $ {\cal T}_{0}$. This is the 
lattice equivalent of the continuum 
two-dimensional representation of four dimensional lattice fields.
Letting $N_{12}=L_1L_2$ and $N_{03}=L_0L_3$, the eigenvalues of 
$  {\cal T}_{1}$ are given by $ \exp\{ \imath \frac{m}{N_{12}} \}  $
and those of   $  {\cal T}_{0}$ are given by $ \exp\{ \imath
    \frac{m'}{N_{03}}\} $, were $m$ ($m'$) are integers defined modulo 
$N_{12}$ ($N_{03}$). An arbitrary vector in the space of lattice 
fields can be expanded in this basis with coefficients $h_{m,m'}$.
From the commutation relations the action of the 
remaining operators $  {\cal T}_{2}$ and  $  {\cal T}_{3}$ 
can be derived by manipulations from the 
previous section applied to each of the $1-2$ and $0-3$ planes.

For the general $q$,$q'$ case, the centralizer of the translation group 
might be non-trivial,  
leading to different subspaces invariant under translations. 
The subspaces are labeled and mapped into each other 
by subgroups  of the continuum group 
generated by the $K^{(\mu)}$. If   
$L_1$ and $L_2$ are divisible by  $q$, and  $L_0$ and $L_3$ are divisible by
$q'$, we obtain $qq'$ subspaces and labellings, just as in the continuum. 
On the opposite extreme, when the lengths $L_{\mu}$ are coprime with the 
associated integers $q$ or $q'$, the subgroup is trivial and we have no 
invariant subspace. In the general case we work with that subgroup of 
the continuum group generated by $K^{(\mu)}$ which maps lattice points into other 
lattice points.

We set now $q=q'=1$ and study 
 the spectrum of four-dimensional lattice Dirac
operators in the constant background field given by the 
links Eq.~(\ref{ConstGF}).
We focus on the upper 
component, $\Psi^+$, because, as in the continuum, 
the lower component can be obtained by charge conjugation. 

 The naive lattice Dirac operator is given by 
\be
D_N=\frac{1}{2} \sum_{\mu} 
\gamma_{\mu} ({\cal T}_{\mu}-{\cal T}^\dagger_{\mu})=\pmatrix{0& \hat{D}_L\cr
\overline{D}_L & 0} 
\ee
where we have defined the lattice Weyl operators in a way similar to the 
continuum. The non-zero spectrum of $D_N$
can be obtained by studying the eigenvalues and eigenvectors of the 
$\hat{D}_L\overline{D}_L$ acting on Weyl 
spinor lattice fields. The spectrum of this 
operator is the same as that of $\overline{D}_L\hat{D}_L$. 
Unlike in the continuum, on the lattice the spectra are the same
even including zero modes: for every positive chirality zero-mode
there is a negative chirality one. 

In section 4 we related the spectrum of the 
4-dimensional Dirac operator to that of the 2-dimensional case and this relation 
extends to the lattice. 
${\cal T}_{1,2}$ commute with ${\cal T}_{0,3}$ and hence 
$\hat{D}_L\overline{D}_L$ becomes block-diagonal.
\be
\hat{D}_L\overline{D}_L=-\pmatrix{D_{12}^\dagger D_{12}+ 
D_{03}^\dagger D_{03}
& 0\cr 0& D_{12}D_{12}^\dagger + D_{03} D_{03}^\dagger }
\ee 
where 
\begin{eqnarray}
D_{12}&=&\frac{1}{2}\left( {\cal T}_1- {\cal T}_1^{\dagger} +\imath\left( 
{\cal T}_2- {\cal T}_2^{\dagger}\right) \right)\\
D_{03}&=&\frac{1}{2}\left( {\cal T}_0- {\cal T}_0^{\dagger} +\imath\left( 
{\cal T}_3-{\cal T}_3^{\dagger}\right) \right)
\end{eqnarray}
$D_{12}$ and $D_{03}$ commute and our two dimensional analysis
applies (see Eq.~(\ref{DOT_Def}). Let $\lambda_{12}$ and 
 $\lambda_{03}$ be the eigenvalues of $D_{12}^\dagger D_{12}$
and  $D_{03}^\dagger D_{03}$  respectively. Then the spectrum of the 
naive lattice Dirac operator is given by $\pm\imath\sqrt{\lambda_{12}+
\lambda_{03}}$ and is two-fold degenerate. For the corresponding 
naive Dirac Hamiltonian $H_N=\gamma_5D_N$, the spectrum is given by 
the same formula without the $\imath$ factor. We emphasize that  
$D_{12}^\dagger D_{12} =H_{NR}^2$, where $H_{NR}$ is the staggered reduced 
Hamiltonian appearing in the last section above equation~(\ref{DOT_Def}). 
Eigenvalues and eigenvectors can be readily constructed in terms 
of those of $H_{NR}$.

The reduction  of the 4-dimensional naive lattice Dirac operator to 
two dimensions does not extend to the Wilson-Dirac or the overlap operator. 
Here, all components are coupled to each other. A complete analysis, 
which will not be included here, can be
done along the lines of the  last section.

\section{Numerical and Asymptotic results}
In this section we will complement the results of the previous
sections by providing some numerical and analytical results on the
spectra of the naive lattice Dirac operator and the overlap Dirac
operator\cite{lattchiral, Neuberger} in a  constant field  
strength gauge background for certain 
values of  $L_{\mu}$ and topological charge. In  particular, we will 
compare the   
eigenvectors of the lattice Dirac operators with their 
continuum  counterparts. The numerical method used for the 
determination of the eigenvalues is an exact diagonalization procedure. 
This method, although precise and stable, is computationally costly, 
limiting what we can do.

We begin by describing our two dimensional results. 
For the numerical determination we  diagonalize the lattice Dirac 
operators (naive and overlap), with  link variables  given by:
\begin{eqnarray}
U_1(n)&=&\exp\{\imath \pi q \frac{I(n_2,n_1)}{L_1 L_2}\}\\
U_2(n)&=&\exp\{-\imath \pi q \frac{I'(n_1,n_2)}{L_1 L_2}\}
\end{eqnarray}
where $I(n_2,n_1)=n_2$ for $1\le n_1 < L_1$ and $I(n_2,L_1)=(L_1+1)n_2$
(and a similar relation for $I'$ exchanging the $1$-$2$ labels).
The value of the plaquette is  constant
and equal to $\exp\{-\imath 2 \pi \frac{q}{N}\}$, with $N=L_1 L_2$.
The associated lattice Dirac operators (matrices) have dimension $2N$. The cases of very
small and very large $N$ can be studied analytically. For $N=2$ the naive
Dirac operator $D_N$ vanishes. For $N=4$ there are again
$4$ zero eigenvalues. In addition, there are two doubly-degenerate pairs at 
$\pm \imath
\sqrt{2}$. For $N=6$ there are three fourfold-degenerate eigenvalues at $0$
and $\pm\imath \sqrt{3/2}$. We have numerically diagonalized 
the naive
lattice Dirac operator $D_N$ for several values of $q$ and $L_i$. In the table 
we display a selection of the eigenvalues for one typical case with $q=1$
and  $L_1=L_2=24$.   We dropped the complex factor $\imath$
from the eigenvalues and list only the positive branch on account of the
spectral symmetry about zero.
Since the continuum eigenvalues are given by 
$\sqrt{2Fn}$,
where $F=2 \pi q/N$ is the magnetic field, 
we  display the square of the eigenvalues divided by $2F$,
a dimensionless quantity.
We see that  the degeneracy of the eigenvalues (indicated inside square
brackets) is equal to $4$.  Only a factor of 2 is
exact at all even values of $N$. The extra twofold degeneracy is approximate, but 
extremely accurate for large values of $N$.

For large $N$ the low lying eigenvalues are well described by:
\be
\label{contlimit}
\imath \lambda_n=\sqrt{2 F n} (1+ C  F n + O(F^2))
\ee
with $C=-1/4$. The dependence on $N$ and $q$ 
only enters through $F$ and we are close
to the continuum limit. $C$ was determined by expanding $D_{12}$ to first 
subleading order in terms of continuum operators:
\be
D_{1 2}=D_1+\imath D_2 + \frac{1}{6}(D_1^3+\imath D_2^3) + \ldots
\ee
In the
one-dimensional quantum mechanical representation we get:
\be
D_{1 2}^\dagger D_{1 2} = 2 F a^\dagger a  -
\frac{F^2}{6}\left(a^4+(a^\dagger)^4+ 6 (a^\dagger a)^2\right) +\ldots
\ee
The eigenvalues of $-D_{1 2}^\dagger D_{1 2}$ are the square of the eigenvalues
of $D_N$, and perturbation theory predicts  
the observed value for $C$ in equation~(\ref{contlimit}).

We have also studied the eigenvalues and eigenvectors of  the overlap operator
$D_O$. $D_O$ is constructed in terms of the unitary operator
$\Omega=\tau_3 H_{WD}(-M,r)/|H_{WD}(-M,r)|$, where $H_{WD}(-M,r)$ is the Wilson
Dirac Hamiltonian with mass $-M$ and Wilson parameter $r$.
The eigenvalues of $\Omega$ are of the form
$e^{\imath \delta_O(M,r)}$ with the $M$ and $r$ dependence explicitly
indicated. Chirality implies symmetric spectrum
and only the positive values will be shown. The continuum eigenvalues
are approached by $M \delta_O(M,r)$ and results are presented
for the square of this quantity divided by $2F$.
In terms of $\delta_O(M,r)$ the eigenvalues of the 
overlap operator are:
\begin{equation}
M(1-\cos\delta_O(M,r) +\imath \sin\delta_O(M,r))
\end{equation}  
This is the operator whose inverse gives the propagator
for internal fermions lines.  External lines should be described 
instead by the exactly chiral invariant operator
$D_I=2\imath M \frac{1+\Omega}{1-\Omega}$~\cite{lattchiral, int-ext}. 
The eigenvalues of the latter are:
\be
\imath 2M\tan(\delta_O(M,r)/2)
\ee

Determining the eigenvalues for small values of $N$ is somewhat more
involved than for the naive operator, since now in addition we must keep
track of the $M$ and $r$ dependence. For example for $N=2$, $q=1$  there is a region
of values of these parameters for which $\Omega$ is equal to $I$ and another
one in which it is equal to $-I$. For $2r^2-4Mr+M^2<0$ (which includes the
point $r=M=1$) $\Omega$ has two eigenvalues equal to $1$ and
two equal to $-1$.
The table contains a selection of eigenvalues
for  $q=1$,   $L_1=L_2=24$,  $r=1$ and $M=0.75,1, 1.25$.
For large $N$, the behavior of the low lying spectrum 
can be understood analytically. 
The eigenvalues are again given by ~(\ref{contlimit}) 
where the constant $C\equiv C_{\rm overlap}(M,r)$ is:
\be
\label{Coverlap}
C_{\rm overlap}(M,r)=-\frac{1}{4}+\frac{r}{M}-\frac{2}{3M^2}
\ee
For $r=1$ and $M=0.75,1,1.25$ $C=-11/108,1/12,37/300$ respectively, 
which works well for the numbers in the table.

\begin{center} 
\begin{tabular}{|c |c | c|c| c|}\hline
Cont. &Naive& Ovp, M=1& Ovp, M=0.75& Ovp, M=1.25\\ \hline
0 & 0 [4] & 0 & 0 & 0 \\ \hline
1 & 0.9946 [4] &  1.0018 & 0.9978& 1.0027  \\ \hline
2 & 1.9783 [4] &   2.0071 &1.9910 & 2.0108\\ \hline
10 & 9.4642 [4]  &  10.167 &9.7723 & 10.265\\ \hline
20 & 17.893 [4] &  20.615& 19.080& 21.040\\ \hline
50 & 37.371 [4] &  52.902& 44.316& 56.123\\ \hline
100& 57.805 [4] &  106.45 & 78.832& 121.57\\ \hline
200 &  & 199.11 & 128.617& 259.15\\\hline
500 & &  350.07 & 202.842& 523.46\\\hline
\end{tabular}
\end{center}

(\ref{Coverlap}) is found by expanding 
the lattice operator in terms of the continuum one. 
We separate the Wilson-Dirac operator into additive terms as follows:
\be
D_{WD}=-M+ D_N +r W 
\ee
$W$ stands for the Wilson term, which is of order $a^2$, while
the naive operator $D_N$ is order $a$. $a$ is the lattice spacing;
equivalently, we set the lattice spacing to unity and use 
powers of $\frac{1}{\sqrt{N}}$ instead.
We write $\Omega=\exp\{-{\cal E}/M\}$, where ${\cal E}$ is an
antihermitian matrix. In the continuum limit we get:
\be
\label{ExprE}
{\cal E}=D_N+\frac{1}{3 M^2} D_N^3 +\frac{r}{2 M}(D_N W+W D_N) + \ldots
\ee
The above holds to order $a^3$. In the continuum limit, 
$W$ can be expressed in terms of the Dirac operator $D$:
\be
\label{ExprW}
W=\frac{-1}{2}(D^2+\imath F_{\mu \nu}\gamma^\mu \gamma^\nu) 
\ee
Expressions~(\ref{ExprE}-\ref{ExprW}) are hold in any dimension
and for arbitrary background. In two-dimensions the anticommutator
of $W$ and $D$ is given by:
\be
\label{WDDW}
WD+DW=-(D^3+ \imath (D_\mu F_{\mu \nu})\gamma^\nu) 
\ee
The second term is proportional to the euclidean equations of motion and
hence vanishes for any classical solution including the uniform
fields we are considering. 
Hence, the behavior of the eigenvalues in the continuum limit 
can be extracted from our previously derived results for 
the naive lattice Dirac operator. In this way we obtain 
Eq.~(\ref{contlimit}) with the $C$ of Eq.~(\ref{Coverlap}).
The second and third term in Eq.~(\ref{ExprE}) are proportional
to $D^3$ and therefore affect the eigenvalues without 
modifying the eigenvectors.

For typical values of the parameters $M,r$ (including $r=M=1$)
the lattice corrections are smaller for the overlap than for the naive
Dirac operator. For example, for $r=1$ and $M=2(1-1/\sqrt{3})$ the leading
$O(a^2)$ correction to the eigenvalues vanishes. This also 
follows from  the numerical data in the table, where the spectrum for
$r=M=1$ is found to be fairly close to continuum for the lowest several 
hundred eigenvalues. 

To conclude the description of the two-dimensional case we will
show  some results on the  low-lying  eigenvectors
of the overlap ($r=M=1$).  In Figs 1 and 2 we display the shape of
the 
positive chirality part of the ground state and the second excited 
state respectively. The data are obtained from the numerical
diagonalization at $L_i=24$. The resulting two-dimensional array of 
points is transformed to the one-dimensional representation and then 
compared to the continuum result, given by eigenstates  of 
the Harmonic oscillator.

\begin{figure}
\begin{center}
\epsfig{file=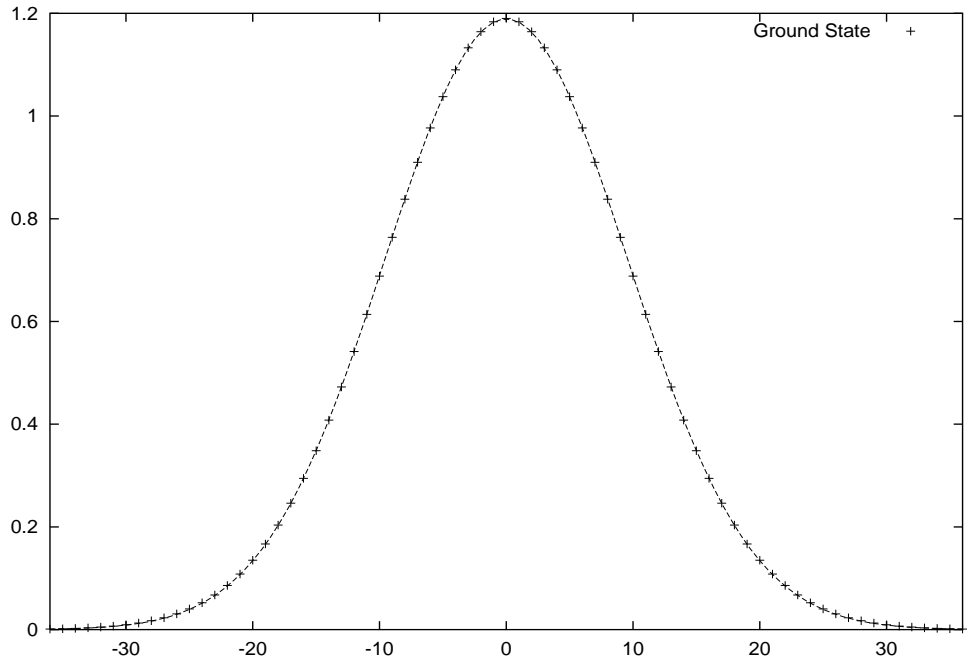,width=13.5cm,height=9cm}
\end{center}

\vskip -0.9cm

\caption{\label{figure 1} The shape of the overlap  (positive
chirality) zero-mode 
in the one-dimensional representation compared to the continuum 
gaussian, for a 2-dimensional constant field with $q=1$ and $L_1=L_2=24$.
}
\end{figure}

\begin{figure}
\begin{center}
\epsfig{file=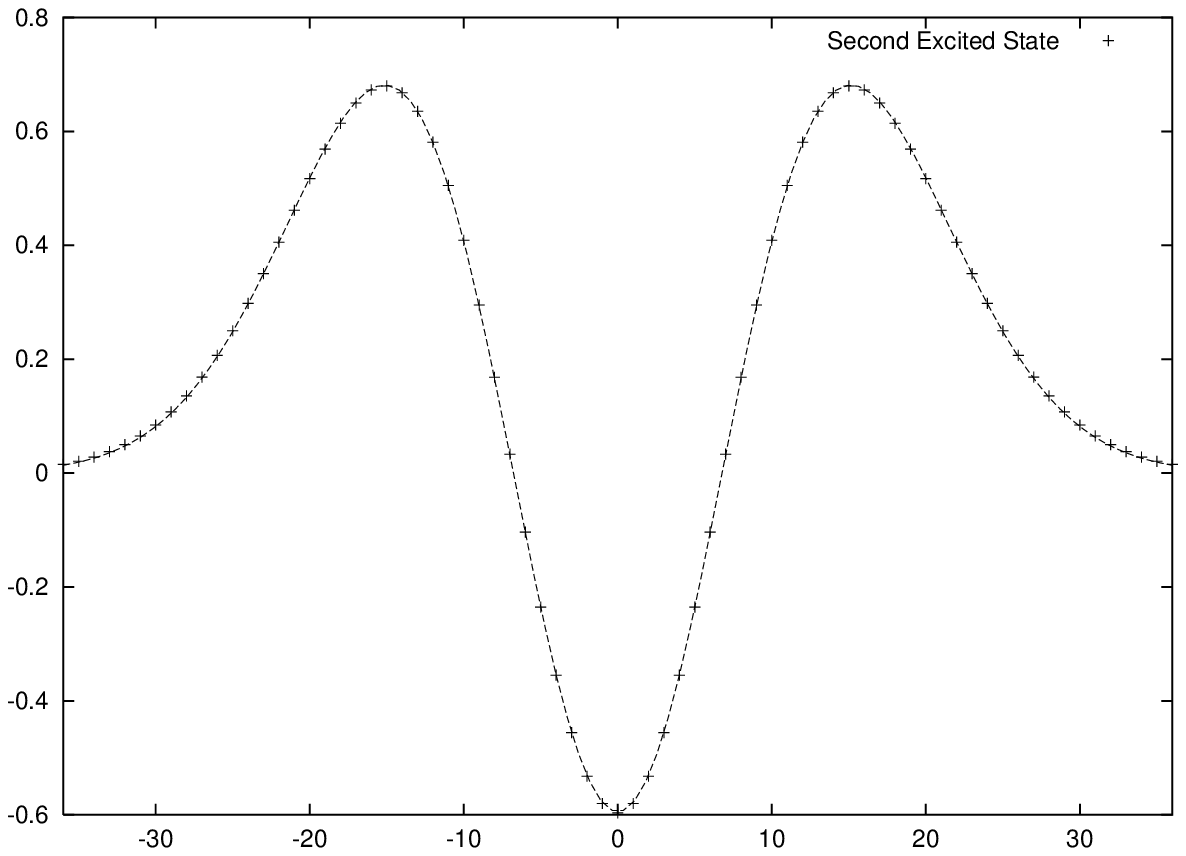,width=13.5cm,height=9cm}
\end{center}

\vskip -0.9cm

\caption{\label{figure2} The shape of the  positive chirality component 
of the second  excited state of the overlap   
compared with the continuum prediction, for  $q=1$ and $L_1=L_2=24$.}
\end{figure}

We turn now to four dimensions. Our numerical work 
is done for $SU$(2) and for gauge fields whose only 
non-zero components are $F_{1 2}=-F_{2 1}= F \tau_3$ and
$F_{0 3}=-F_{3 0}= F' \tau_3$. The color components decouple
so we essentially are dealing with $U(1)$ systems.
As explained in section 6, the eigenvalue problem for the naive Dirac operator 
reduces to the two-dimensional problem. Here we focus on the 
overlap operator, which does not simplify similarly. 

Our objective is to find the leading corrections to 
the low-lying spectrum when continuum is approached. 
Eqs.~(\ref{ExprE})-(\ref{ExprW}) are valid now, but
Eq.~(\ref{WDDW}) is replaced by:
\be
WD+DW=-(D^3+ \imath (D_\mu F_{\mu \nu})\gamma^\nu+\imath \epsilon^{\alpha
\rho \mu \nu} \gamma^5 \gamma^\alpha  F_{\mu \nu}  D_\rho ) 
\ee
where we have made use of the Bianchi identities.
The second term vanishes for gauge configurations that solve
the classical equations of motion.
The last term was absent in two-dimensions. For a uniform field this
term commutes with $D$. All the terms in ${\cal E}$
anticommute with $\gamma_5$ and are antihermitian, so that
in the chiral basis ${\cal E}$ has the form:
\be
{\cal E}=\pmatrix{0&-X^\dagger\cr X &0}
\ee
with $X$ given by:
\begin{eqnarray}
X=\bar{D}_N+(\frac{1}{3 M^2}-\frac{r}{2
M})\bar{D}\hat{D}\bar{D} \nonumber \\
-\frac{r}{ M}(F(D_0+\imath \tau_3 D_3)+F'(\imath
\tau_1 D_1 +\imath \tau_2 D_2))\tau_3+\ldots
\end{eqnarray}
The eigenvalues of $X^\dagger X$ are obtained 
by writing the covariant derivatives in terms of creation and annihilation
operators and using perturbation theory. With our choice
of gamma matrices $X^\dagger X$ is already diagonal in spin space.
The $1-1$ component of  $X^\dagger X$ is readily expressed in terms of
the two number operators. The corresponding eigenvalue of  $X^\dagger X$
becomes:
\begin{eqnarray}
\label{eigenXX}
2 F n_{12} +2 F' n_{0 3}
+4(-\frac{1}{4}+\frac{r}{M}-\frac{2}{3M^2})(F n_{12} + F' n_{0
3})^2 \nonumber \\
-\frac{4r}{M}(F^2n_{12} + F'^2 n_{0 3})
\end{eqnarray}
where $n_{12},n_{03}$ are non-negative integers. 
The first two terms give the continuum result, the third
one is of a type familiar from the two-dimensional case, while the
fourth one is genuinely four-dimensional. The $2-2$ spin-component of
$X^\dagger X$ gives another set of eigenvalues given by a formula 
identical to Eq.~(\ref{eigenXX}) except that the last term has the
opposite sign and $n_{12},n_{03}$ are now integers strictly larger than zero.
The sign change in the last term lifts the degeneracy of the continuum 
eigenvalues. 
The eigenvalues of  
${\cal E}$ are
given by those of $\pm\imath \sqrt{X^\dagger X}$.

We have numerically determined the eigenvalues
of the overlap operator in four-dimensions. Our exact
diagonalization methods cannot be carried out for very large sizes and a
comparison with the previously derived formulas is difficult.

\section{Summary}

With one exception, any $U(1)$ or $SU(N)$ gauge field on a two dimensional
or four dimensional Euclidean torus can be smoothly deformed to an abelian background.
When the background
is topologically nontrivial the abelian field can be deformed to a non-vanishing
uniform magnetic
field. Any topological invariant is therefore captured by a representative with uniform
abelian background. The Dirac equation in these backgrounds
can be viewed as an equation over infinite space but in half the dimension. This naturally
leads to an exact diagonalization in terms of harmonic oscillator wave functions. 
In the massless case one gets explicit formulas for the expected zero modes.
Much of this translates to the lattice; for example the diagonalization of the Wilson
Dirac operator and the associated overlap Dirac operator on a two dimensional torus
simplifies to an equation on a one dimensional circle of length fixed by the area
of the original torus. In these backgrounds the connection between fermions and topology
becomes particularly transparent both in the continuum and on the lattice, 
once one employs overlap fermions there.

\section{Acknowledgments}

Most of this work was developed during a six month stay of A.~G-A at Boston 
University (BU). A.~G-A wishes to thank the Physics Department, and especially 
the high energy theory group at BU for their hospitality. A.~G-A acknowledges 
financial support from Spanish Ministry of Education and from grant FPA2000-0980. 
The research of L.~G., Ch.~H. and C.~R. was supported in part under 
DOE grant DE-FG02-91ER40676.
The research of H.N. was supported in part by DOE grant DE-FG02-96ER40949
and by a Guggenheim fellowship. H.~N.~would like to thank L.~Baulieu and the
entire group at LPTHE for their hospitality and support.


\newpage 

\begin{thebibliography}{99}

\bibitem{Landau} L. Landau, Z. f{\" u}r Physik 64 (1930) 629. 

\bibitem{abrikosov} A.A. Abrikosov JETPP 32 (1957) 1442. Reprinted in 
{\em Solitons and Particles} ed. by C. Rebbi and G. Soliani,
World Scientific (1984).

\bibitem{saviddy}
M.J. Duff and M.  Ram\'on-Medrano, Phys. Rev. D12 (1975) 376;
I.A. Batalin, S.G. Matinyan and G.K. Saviddy, Yad. Fiz. 26 (1977) 407;
S.G. Matinyan and G.K. Saviddy, Nucl. Phys. B134 (1978) 539;
G.K. Saviddy, Phys. Lett. 71B (1977) 133;
H. Pagels and E. Tomboulis, Nucl. Phys. B143 (1978) 485;
N.K. Nielsen and P. Olesen, Nucl. Phys. B144 (1978) 376;
N.K. Nielsen and P. Olesen, Phys. Lett. 79B (1978) 304.

\bibitem{leutwyler}
H. Leutwyler, Phys. Lett. 96B (1980) 154;
Nucl. Phys. B179 (1981) 129.      

\bibitem{thooft}
G. 't Hooft,  Commun. Math. Phys. 81 (1981) 267.

\bibitem{vanbaal2} P. van Baal, Commun. Math. Phys. 94 (1984) 397.

\bibitem{vanbaal1}
P. van Baal,  Commun. Math. Phys. 85 (1982) 529.

\bibitem{smitvink} J. Smit and J. Vink, Nucl. Phys. B286 (1987) 485.

\bibitem{domain}D.~B.~Kaplan, Phys.~Lett.~B288 (1992) 342.

\bibitem{lattchiral} R. Narayanan and H. Neuberger, Phys. Lett. B302 (1993) 62;
Phys. Rev. Lett. 71 (1993) 3251; Nucl. Phys. B443(1995) 305. H. Neuberger,
Phys. Rev. D57 (1998) 5417; ``Exact chiral symmetry on the lattice'', to
appear in Annu. Rev. Nucl. Phys. Part. Sci. 2001, Vol 51. 

\bibitem{gattringer1} C.R. Gattringer, I. Hip and C.B. Lang, 
Nucl. Phys. B508 (1997) 329 (hep-lat/9707011).

\bibitem{chiu} Ting-Wai Chiu, Phys. Rev. D58 (1998) 074511
(hep-lat/9804016).

\bibitem{gattringer2} C. Gattringer and I. Hip, Nucl. Phys. B536 
(1998) 363 (hep-lat/9712015).

\bibitem{gpgap}
M. Garc\'{\i}a Perez, A. Gonz\'alez-Arroyo and C. Pena, JHEP 0009 (2000) 033.


\bibitem{douglas}
 M.R. Douglas and N.A. Nekrasov, {\em Non-commutative Field Theory} 
hep-th/0107110.

\bibitem{tek} A. Gonz\'alez-Arroyo and M. Okawa Phys. Lett. 120B (1983) 174;
Phys. Rev. D27 (1983) 2397;
T. Eguchi and R. Nakayama, Phys. Lett. 122B (1983) 59; 
A. Gonz\'alez-Arroyo and C.P. Korthals Altes, Phys. Lett. 131B (1983) 396

\bibitem{sushel} L. Susskind,  hep-th/0101029;
S. Hellerman and M. Van Raamsdonk, JHEP 0110 (2001) 039. 

\bibitem{te}  J. Groeneveld, J. Jurkiewicz and C. P. Korthals Altes, Phys.
Scr. 23 (1981) 1022; J. Ambjorn and H. Flyvberg, Phys. Lett. 97B (1980) 241;
B. van Geemen and P. van Baal, J. Math. Phys. 27 (1986) 455; D.R. Lebedev and M.I. Polikarpov,
Nucl. Phys. B269 (1986) 285.


\bibitem{review}
 A. Gonz\'alez-Arroyo, {Gauge fields on the  four-dimensional torus} 
 Proceedings of the Pe\~niscola 1997 Advanced School on Non-Perturbative Quantum Field 
 Physics, World Scientific   (1998). 

\bibitem{Neuberger} H. Neuberger, Phys. Lett. B417 (1998) 141;
Phys. Lett. B427 (1998) 353.

\bibitem{int-ext} H. Neuberger, Nucl. Phys. Proc. Suppl. 73 (1999) 697.

\end{thebibliography}
\end{document}